\documentclass{evn2004}
\usepackage{txfonts}
\usepackage{graphicx}
\begin{document}
   \title{Compact Structure in FIRST Survey Sources} 

   \author{R. W. Porcas\inst{1},
           W. Alef\inst{1},
           T. Ghosh\inst{2},
           C. J. Salter\inst{2}
          \and
          S. T.  Garrington\inst{3}}

    \institute{Max-Planck-Instit\"ut f\"ur Radioastronomie, Auf dem H\"ugel 69, D-53121 Bonn, Germany
         \and
     Arecibo Observatory, HC3 Box 53995, Arecibo, PR 00612, USA
         \and
     Jodrell Bank Observatory, University of Manchester, Macclesfield, Cheshire, SK11~9DL, UK}

   \abstract{
We present preliminary results from a statistical survey of compact
structure in faint radio sources. Around 1000 sources from the VLA
FIRST survey (flux densities $\geq$ 1~mJy at 1.4~GHz) have been observed
with the single-baseline interferometer Effelsberg-Arecibo. We
observed each source, selected from a narrow strip of sky at
declination 28$^\circ$, for just one minute. The baseline sensitivity at
1.4~GHz, using 512~Mb/s recording, is such that {\it any} FIRST source,
selected at random, would be detected if most of its flux density is
in compact structure. We discuss the detection-rate statistics from
one epoch of these observations.
   }

   \maketitle

\section{Introduction}
We have been investigating the existence of compact structure in the
faint radio source population using a single, but highly sensitive,
long-baseline interferometer between the \mbox{305-m} Arecibo and 100-m~Effelsberg
radio telescopes. Preston et al. (1985) showed the value of
such observations by surveying 1398 known radio sources at 2.29~GHz between 1974 and 1983,
using single baselines between the (then) 64-m~DSN antennas, in order
to select candidate sources for use in establishing the celestial reference frame.
Sparse u,v-coverage VLBI observations were also used by Lawrence et al. (1985) at
22~GHz to investigate highly compact structure in radio sources, as a preliminary
to future space-VLBI projects.

In the same spirit as these previous surveys, we decided to investigate the
prevalence of compact structure in a sample of sources chosen from the 1.4~GHz FIRST
survey (Becker et al., 1995).
This survey, made using the VLA with a resolution of 5.4'',
contains 138,665 sources down to 1~mJy in the first instalment (White et al., 1997),
with a sensitivity of 0.15~mJy.
Garrington et al. (1999) have made Global 5-GHz observations of a small sample
of 35 FIRST sources, with flux densities $\geq$~10~mJy and preselected using
MERLIN to have compact structure.
They detected 27 sources, using the phase-referencing technique.

In our investigation we have used a highly sensitive
long-baseline interferometer, between Arecibo and Effelsberg,
using the FIRST survey frequency of 1.4~GHz and recording using 512~Mb/s.
For a given integration time
this baseline is 42~times more sensitive than 
a single VLBA baseline, 6.2~times more sensitive than the full
(phase-referenced) VLBA, and 12.5~times more sensitive than
the full VLBA at its sustainable recording rate of 128~Mb/s.
We believe this is the most sensitive VLBI baseline ever used.

\section{Observations and Source Selection}

We made VLBI observations between Arecibo and Effelsberg
using MK4 recording at 512~Mb/s.
We observed the 64-MHz band from 1.366-1.430~GHz in both LHC and
RHC polarization, using 2-bit sampling, and subdivided into
4 upper and 4 lower sideband channels, each of 8~MHz,
for each polarization.
At Effelsberg the signal was recorded using a MK5A system. At
Arecibo the data was recorded on a VLBA4 recorder using 2 heads.
We configured this so that the LHC signal was recorded by one
head, and RHC by the other, thus making it convenient to
make a preliminary analysis of the polarizations separately from the results of the
necessary 2-pass correlation (see Section~3).

Since the Arecibo telescope has a limited hour-angle range, we decided to
make ``single-shot''
observations of FIRST sources at a fixed hour-angle. We observed sources
in a 1$^\circ$-wide declination strip between declinations 28$^\circ$ and 29$^\circ$
at $\sim$30~mins before
Arecibo transit, in order to minimize Arecibo drive times between sources
whilst being at a reasonable elevation ($\sim$34$^\circ$) at Effelsberg.
The projected baseline results in a resolution of $\sim$6~mas in
PA~$\sim$20$^\circ$
We planned an on-source integration time of 60~s, for which the estimated
rms noise is 0.1~mJy.
This permits us to detect {\it any} FIRST source if it is sufficiently
compact.
We allowed 30~s for telescope drive-time between sources, resulting in
observation of 15 target sources for every 22~min tape pass.
We initially planned four 6~hour observing periods, separated by a few
weeks; in the end we had 4 observing epochs in October 2003, March and June 2004,
providing a total of 27.5 hours of observing.
At our first observing epoch we were also able to add the 76-m~Jodrell
Bank Lovell Telescope and the phased Westerbork Array,
for the purpose of investigating short-baseline flux-densities
for some of the stronger FIRST sources.

From the outset we decided to make an un-biased selection from the
FIRST catalogue, without regard to flux density, size or radio
spectrum. Our motivation is not to identify ``compact radio sources''
but, rather, to make a $\it statistical$ survey of
the existence of compact structure in $\it all$
source types.
We thus calculated a notional right ascension for each 60~s observation
and selected the first source in the FIRST catalogue with an RA 
greater than this within our declination strip, provided it had not
been observed in a previous epoch.
We also identified 11 VLBA calibrator sources within the strip. At each epoch these were
substituted for the closest target sources selected.

There are 11,699 sources listed in White et al. (1997) within our
declination strip. Excluding the calibrator sources, we scheduled
992 target sources (215+259+259+259) for our 4 observing epochs.
However, there is often more than one FIRST source within 
the Arecibo beam (FWHM = 210'') and we are able to investigate
a further $\sim$400 FIRST sources using multi-pass correlation.

\section{Correlation and Fringe Detection}

We are correlating these observations using the MK4 correlator at the MPIfR in
Bonn. As the tape-drives are equipped with only a single playback
head it is necessary to make at least 2 correlation passes, one for each head (polarization).
Further passes are needed to correlate other FIRST sources which are known to
be within the same Arecibo beam of the target sources. Their response is
necessarily attenuated by up to 0.7; note that as the Effelsberg FWHM beam
width is much larger it is the Arecibo amplitude beam which is relevent.
We use a correlator mode providing a pre-average time of 1~s and 128 delay
steps per 8~MHz sideband (equivalent to 62.5~kHz frequency resolution).
Correlation is inhibited under control of Efflesberg and Arecibo 
off-source flags.
For most sources, between 60 and 70~s of correlated data are recovered.

We use the Haystack HOPS package fringe-fitting program FOURFIT to search
for a source response in the correlator output for each source.
Source positions from the FIRST catalogue have accuracies of $\sim$1''
or better. The delay and fringe-rate resolution of our observations
are typically 170~mas in PA~20$^\circ$ and 2'' in PA~--50$^\circ$,
respectively.
We can search using a range of delay and rate windows to optimise
our detection threshold whilst taking into account possible instrumental
and FIRST source position errors.

 \begin{figure*}
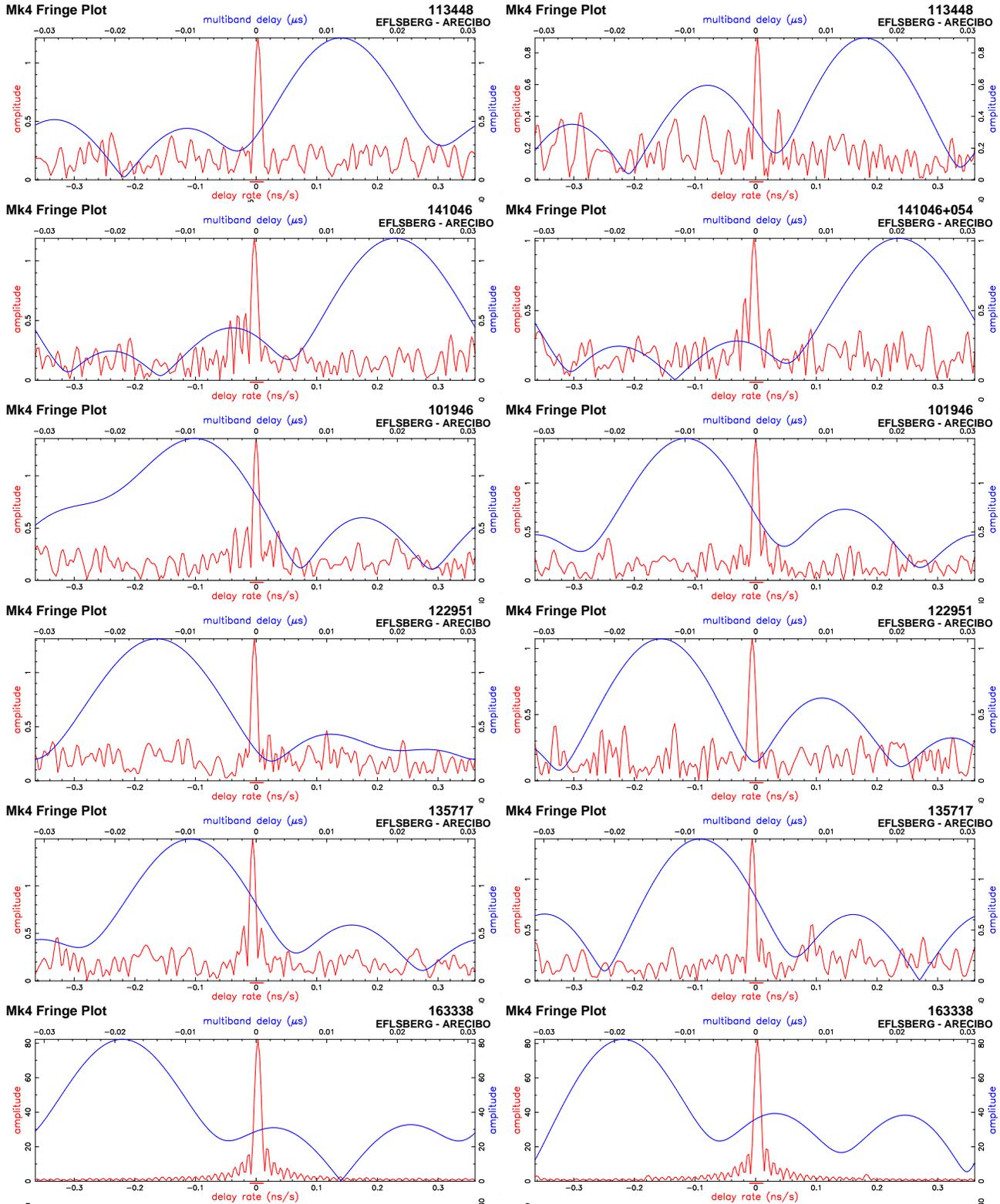

   \centering
   \includegraphics[bb=25 579 499 769,width=0.97\columnwidth,clip]{RPorcas_1_fig1a.epsi}
   \includegraphics[bb=25 579 499 769,width=0.97\columnwidth,clip]{RPorcas_1_fig1b.epsi}
   \includegraphics[bb=25 579 499 769,width=0.97\columnwidth,clip]{RPorcas_1_fig1c.epsi}
   \includegraphics[bb=25 579 499 769,width=0.97\columnwidth,clip]{RPorcas_1_fig1d.epsi}
   \includegraphics[bb=25 579 499 769,width=0.97\columnwidth,clip]{RPorcas_1_fig1e.epsi}
   \includegraphics[bb=25 579 499 769,width=0.97\columnwidth,clip]{RPorcas_1_fig1f.epsi}
   \includegraphics[bb=25 579 499 769,width=0.97\columnwidth,clip]{RPorcas_1_fig1g.epsi}
   \includegraphics[bb=25 579 499 769,width=0.97\columnwidth,clip]{RPorcas_1_fig1h.epsi}
   \includegraphics[bb=25 579 499 769,width=0.97\columnwidth,clip]{RPorcas_1_fig1i.epsi}
   \includegraphics[bb=25 579 499 769,width=0.97\columnwidth,clip]{RPorcas_1_fig1j.epsi}
   \includegraphics[bb=25 579 499 769,width=0.97\columnwidth,clip]{RPorcas_1_fig1k.epsi}
   \includegraphics[bb=25 579 499 769,width=0.97\columnwidth,clip]{RPorcas_1_fig1l.epsi}
   \caption{
    Fringe plots of some of the sources detected, in RHC (left) and LHC (right)
    polarization.
    In each panel the jagged plot is the residual fringe-rate spectrum; the
    smoother plot is the multiband-delay function.
    The top 5 rows are of the 5 detected sources with lowest
    listed flux densities in the FIRST catalogue (1.03,
    1.11, 1.21, 1.27 and 1.28~mJy).
    The bottom row is for the source with the highest S/N-ratio detection (620-sigma).
}
    \end{figure*}
%

\section{Preliminary Statistical Analysis}

We present here a very preliminary analysis of results from our
second epoch, observed on 22~March 2004 between UT~02h00 and 09h11.
It represents only about a quarter of our sample, and uses a separate
search in the LHC and RHC data, using conservative (wide) search
windows and uncalibrated data.
Some 7 of our 259 targets were lost due to a recording
malfunction, resulting in a sub-sample of 252 sources.

A total of 71 (28\%) of the 252 target sources were detected in either LHC or RHC correlations
above a conservative threshold of 8-sigma,
and at the expected residual fringe-rate and delay.
Of these, 63 were detected in both polarizations. Note that the addition
of the polarizations would result in detections above 11-sigma for
these sources. For the sources detected in a single polarization the
response in the other polarization was above 6.8-sigma in all cases.
Analysis of the distribution of S/N-ratio for the detected sources
suggests that at least a futher 14 sources may be detected above 8-sigma
in polarization-added data, bringing the detection rate to 33\%.

We have also searched data from 86 other FIRST sources located within the
target source beams. Some 11 sources were detected, in both polarizations.
This lower fraction (13\%) reflects signal attenuation away from the centre of
the Arecibo beam.

We have investigated the detection rate as a function of the (peak)
flux density listed in the FIRST catalogue.
The parent distribution of flux densities in the
FIRST catalogue (derived from our subsample of 1392 target or in-beam
sources) has quartiles of 1.35, 2.24 and 5.42~mJy, and for the subsample of
252 sources observed here the values are similar: 1.41, 2.24 and 4.86~mJy.
The numbers of sources detected in these ranges are 8, 19, 22 and
22. The median FIRST catalogue peak flux density of our detected
sources is 3.37~mJy. 11 sources are in the range 1.0 to 1.5~mJy.
Fringe detection plots for the 5 detections with lowest listed
flux densities are presented in Figure~1.

\section{Conclusions}

It is, of course, premature to draw any definitive conclusions, but
our preliminary analysis suggests that up to a third of all FIRST
sources contain detectable compact structure at the mJy level.
Note that long, thin structures - e.g. jets - may be missed in our observations
if the structure is oriented along our resolution direction.
Interestingly,
the detection rate for sources $\geq$4.9~mJy is not significantly higher
than that for weaker sources.
Our 8-sigma single-polarization detection threshold corresponds to $\sim$1.1~mJy.
For sources with true flux densities
in the 1-2~mJy range, a detection implies that at least half of the
total flux must reside in compact components. The fall-off in detection
rate between 1.41 and 1.00~mJy may represent this fraction falling
below our detection threshold, or may reflect source variability or increasingly large
errors in the FIRST flux densities at the bottom of the survey.
Once the data have been calibrated, a detailed investigation of the distribution
of source visibilities will allow these effects to be investigated further.

\begin{acknowledgements}
We thank Dave Graham for assistance with observing at Effelsberg,
Emmanuel Momjian for assistance at Arecibo,
and 2 generations of Effelsberg and Arecibo schedulers, Rolf
Schwartz, Alex Kraus, John Harmon and Hector Hernandez, for their
patience in supporting this project.
\end{acknowledgements}

\cleardoublepage


\begin{thebibliography}{}

  \bibitem[1995]{becker} Becker, R.H., White, R.L. \& Helfand, D.J. 1995,
       Ap.J. 450, 559

  \bibitem[1999]{garrington} Garrington, S. T., Garrett, M. A. \& Polatidis, A. 1999,
       New A.R. 43, 629

  \bibitem[1985]{Lawrence} Lawrence, C. R., Readhead, A. C. S., Linfield, R. P.,
     Payne, D. G., Preston, R. A., Schilizzi, R. T., Porcas, R. W., Booth, R. S. \&
     Burke, B. F. 1985,
       Ap.J. 296, 458

  \bibitem[1985]{preston} Preston, R.A., Morabito, D. D., Williams, J.G.,
       Faulkner, J., Jauncey, D. L. \& Nicolson, G. 1985,
       A.J. 90, 1599

  \bibitem[1997]{white} White, R.L., Becker, R.H., Helfand, D. J. \& Gregg, M.D. 1997,
       Ap.J. 475, 479

\end{thebibliography}
\end{document}